\documentstyle[sprocl,psfig]{article}
\def\Journal#1#2#3#4{{#1} {\bf #2}, #3 (#4)}

\def\PRD{{\em Phys. Rev.} D}

\def\be{\begin{equation}}
\def\ee{\end{equation}}
\def\bea{\begin{eqnarray}}
\def\eea{\end{eqnarray}}
\newcommand{\beqn}{\begin{eqnarray*}}
\newcommand{\eeqn}{\end{eqnarray*}}
\newcommand{\dps}{\displaystyle}
\def\msun{M_\odot}
\def\op{ \ $ }
\def\cl{$ \ }

\def\lappreq{\! \stackrel{\scriptscriptstyle <}{\scriptscriptstyle
\sim}\!}

\begin{document}

\title{
WHAT  GRAVITATIONAL SIGNALS SAY ABOUT THE STRUCTURE AND THE EVOLUTION OF ASTROPHYSICAL
SOURCES}

\author{V. FERRARI}
\address{Dipartimento di Fisica ``G.Marconi",
 Universit\` a di Roma ``La Sapienza"\\
 and Sezione INFN  ROMA1, p.le A.  Moro
 5, I-00185 Roma, Italy\\E-mail: valeria@roma1.infn.it}


\maketitle\abstracts{ 
Gravitational waves transport very detailed information on the structure and evolution of 
astrophysical sources. For instance a binary system in the early stages of its evolution
emits a  wavetrain at specific frequencies that depend on the characteristics 
of the obital motion; as the orbit shrinks and circularize, due to radiation reaction
effects, the orbiting bodies get closer and tidally interact. 
This interaction may result in the
excitation of the proper modes of oscillation of the stars, and in the emission of 
gravitational signals that carry information on the mode
frequencies, and consequently   on the equation of state in the stellar interior.
These phenomena may occur either in solar type stars with orbiting planets
and in compact binaries, and in this lecture we will discuss different approaches that can
be used to study these processes in the framework of General Relativity.
}

In recent years,  major progresses have been done in the construction of large 
ground-based interferometers, which will make accessible to observations a 
large frequency region, ranging from a few Hz to a few kHZ:
TAMA has already performed some  observational runs,
reaching a sensitivity that would enable to detect the coalescence of a 
neutron star-neutron (NS-NS) binary
system occuring in our Galaxy with a signal to noise ratio higher than 30, and
VIRGO, LIGO  and GEO600, are almost completely assembled and within the end of the
year 2002 are expected to start scientific runs  \cite{web}.
Resonant bar detectors, ALLEGRO, AURIGA, EXPLORER, NAUTILUS, NIOBE, have been
operational since many years, and are sensitive to a small
frequency region of about 10-20 Hz, centered at $\sim$1 kHz. As an example,
Explorer and Nautilus would be able to detect, with signal to noise
ratio equal to 1,  a burst of gravitational waves  of amplitude $h\sim 4\cdot 10^{-19}$, 
which would correspond to a mass of a few units in $10^{-3}~M_\odot$ 
trasforming into gravitational waves at the center of our Galaxy \cite{astone}.
And finally, the space-based intereferometer LISA, which is expected to fly in about
a decade, will enlarge the observational window 
to the low frequency region $10^{-4}~Hz\lappreq \nu \lappreq
10^{-1}~Hz$.

Many are the astrophysical sources that, according to
the theory of General Relativity, emit gravitational waves in the frequency region spanned  
by these detectors, and in this 
review I will discuss the characteristic properties
of the signals emitted by some of the more interesting sources, and  show   
how to compute them.
The detectability of a signal depends also on the {\it event rate},
namely on how many events  per year occur in the 
volume accessible to observations, and  on the {\it detection rate},
which depends on the responce of each detector to a specific source; 
how these rates are determined is,  of course, a very important issue,
but a discussion of these problems is beyond
the scope of this review, which will be focussed on the information that the gravitational
signals carry on the structure (either geometrical or internal) of the source, on its
motion, and on physical processes that may occur in some particular situations, 
like the resonant excitation of the
stellar modes.

In Sec. I, I will shortly describe the quadrupole formalism, which is the easiest method
to  estimate  gravitational fluxes and waveforms, 
and I will discuss its domain of applicability.
In Sec. II and III, the quadrupole formalism will be applied to rotating neutron stars and
to binary systems far from coalescence, respectively.
In Sec. IV the resonant excitation of the g-modes in extrasolar planetary systems will be
considered in the framework of a perturbative approach, and in Sec. V the gravitational
emission of coalescing binary systems  will be studied.


\section{The quadrupole formalism}\label{sec1}
The quadrupole formalism is one of the most powerful tools to
estimate the amount of radiation emitted by a dynamically evolving system.
It is based on the assumptions that  
the gravitational field is weak, so that gravitational
interaction do not dominate, and that the velocities of the bodies involved in the
problem are much smaller than the speed of light; 
as a consequence, the spacetime metric can be written as the metric of the flat spacetime, $
\eta_{\mu\nu},$ plus a small perturbation
\[
g_{\mu\nu}=\eta_{\mu\nu}+ h_{\mu\nu},\qquad\quad \vert h_{\mu\nu} \vert << 1,
\]
and the Einstein equations can be linearized and solved.
By a suitable choice of the gauge,
the perturbation $ h_{\mu\nu} $ is found to satisfy a wave equation, and to  
propagate in vacuum at the speed of light;
its amplitude is shown to depend only on
the time variation of the energy density of the source, $\rho(t,\bf{x}),$ as follows
\be
\label{quad}
\cases{
\dps{
h^{ij}= \frac{2G}{c^4}
\frac{e^{i\omega\frac{r}{c}}}{r}\left[
\frac{1}{c^2}\frac{\partial^2}{\partial t^2}~ q^{ij}\right]}&\cr
\dps{h^{\mu 0}=0,\quad \mu=0,3
}
&\cr},\qquad
q^{ij}=\int_V \rho(t,{\bf x}) c^2 x^i x^j dV,
\ee
where \op\dps{q^{ij}}\cl  is
the quadrupole moment,  and the integral extends over the
source volume.

An interesting point to stress:
the asumption
\[
v << c \quad\rightarrow\quad  \Omega R << c,\]
where $R$ is the source extension,
implies that the region where the source is confined  must be small compared to
the wavelenght of the emitted radiation,
$\lambda_{GW} = \frac{2\pi c}{\Omega}$.
When this constraint is not verified, the quadrupole approach is no longer
applicable. 
For instance,  the binary system  PSR 1931+16 \cite{hulsetaylor}, whose emission
properties will be
described in some more detail in the next section,
is composed of two neutron stars with masses $M_1\sim M_2\sim 1.4 \msun$, moving 
on an eccentric orbit of diameter $D\sim13.8\cdot 10^{10} ~cm;$
due to the orbital motion it emits gravitational waves at some specific frequencies,
and the highest spectral line is at  $\nu_{GW}\sim 10^{-4}~Hz.$ It follows that
\[
\lambda_{GW}=\frac{c}{\dps{\nu}_{GW}}\sim 10^{14}~cm,
\qquad  \lambda_{GW} >> D,
\]
and the quadrupole constraint is satisfied.
But if we want to describe a pulsating neutron star,  since the frequency of the
fundamental mode of oscillation ranges within $\sim 2-3$ kHz, depending on the
equation of state (EOS), and since the  typical diameter of a NS is $D\sim 20$km,
\[
\lambda_{GW}=\frac{c}{\nu_{GW}}\sim 
\frac{3\cdot 10^{10}~cm/s}{2-3\cdot 10^{3}~Hz}{{ \sim 10^7~cm}},
\qquad\hbox{i.e.}\qquad  \lambda_{GW} \sim D,
\]
and the quadrupole formalism cannot be applied.
To study these problems,  alternative and more sophisticated approaches have to be used
as we shall see in Sec. \ref{sec4}.

\section{Rotating neutron stars}\label{sec2}
As a first application of the quadrupole formalism, we shall consider the case of
rotating neutron stars. 
Neutron stars can radiate their positive rotational energy
essentially in two ways.  If the star is triaxial, it has
a time varying  quadrupole moment and, according to Eq.~(\ref{quad}), emits
gravitational waves  at twice the rotation frequency, and with an amplitude
which can be parametrized in the following way \cite{bongou1}
\be
h \sim 4.2\times 10^{-24}\left(\frac{ms}{P}\right)^2
~\left( \frac{r}{10~kpc}   \right)^{-1}
\frac{I}{10^{45}~g ~ cm^2}~ \left(\frac{\epsilon}{10^{-6}}\right)
\ee
where \op I=10^{45}~g ~ cm^2,\cl is the typical value of the moment of inertia
of a neutron star, $P$ is the rotation period, and
$\epsilon$ is the oblateness of the star, here given in
units of \op 10^{-6}.\cl  
These waves could be detected, for instance, by VIRGO with one year of integration,
if the amplitude of the signal were of the order of
$h \sim 10^{-26}$; this means that neutron stars outside our Galaxy
would practically be undetectable. In addition, much depends
on the value of the oblateness, and
several studies have tried to set limits on its possible range of variation.
Gourghoulon and Bonazzola made a first attempt to set constraints on
$\epsilon$  using observational data \cite{bongou2}.
They considered a  number of pulsars,
and assuming that the observed slowing down of
the period is entirely due to the emission of GW, gave an estimate
of $\epsilon.$ 
They found  \op \epsilon \in \sim [10^{-2}, 10^{-9}],\cl
but of course this can only be an upper bound, since
we know  that rotational energy is dissipated in pulsars also by other mechanisms, like
the electromagnetic emission and/or  the acceleration of charged
particles in the magnetosphere.
Further studies on this problem \cite{ushomirskycutlerbildsten}
established that if the triaxial shape  is due to strains
in the crust of the neutron star, the oblateness
could be \op \epsilon \lappreq 10^{-7}\left( \frac{\sigma}{10^{-2}} \right),
\cl where \op \sigma\cl is the strain needed to break the crust,
but its value is quite  uncertain; for instance
$\sigma \in [10^{-2}, 10^{-1}]$ according to
\cite{smoluchowski}, and $\sigma \in [10^{-4}, 10^{-3}] $
according to  \cite{ruderman}.
A time dependent quadrupole moment can also be due to a precession of the
star's angular velocity around the symmetry axes.
In this case the radiation is emitted at a frequency
\op
\nu_{prec}=\frac{1}{2\pi}(\omega_{rot}+\omega_{prec})\simeq
\nu_{rot},
\cl
but the amplitude of the
precessional contribution depends on   a further parameter,
the ``wobble-angle" between the rotation and the symmetry axes, which is largely
unknown.
From the above discussion, we can conclude that unless astronomical observations
will  set more stringent constraints on the oblateness and
on the wobble-angle of rotating neutron stars, it is difficult, at present, to
say whether the radiation emitted by rotating neutron stars 
will be detectable  in the near future.

It is worth mentioning that the detection of  gravitational waves depends on many 
cooperating factors;
indeed a signal is detectable if
its frequency is in the range of the detector's bandwith, if
its amplitude is high enough to be extracted from the noise
by a suitable filtering technique, and, last but not least, if  the
computing resources are  sufficient to execute this operation.
For instance, in the case of rotating neutron stars,
the estimated number of sources  in the Galaxy is
$\sim 10^9,$ of which, $\sim 1000$ are observed as pulsars, 
and  5 are at a distance smaller that 200pc;
thus, at least the number of possible sources is encouraging; 
for some of them we have some further information, like the location in sky 
and the spin-down rate. However,
we do not know the oblateness and since the star is certainly
in relative motion with respect to the detector, the filters to be used 
to extract the signal from noise must contain the Doppler correction;
thus, to detect  signals from stars of unknown position and characteristics,
we should divide the sky in patches as small as possible, and search for possible
sources by using filters with different parameters, and this is  extremely
expensive!

\section{Binary systems far from coalescence: the quadrupole approach}\label{sec3}
The dynamical evolution of a binary system is affected by the 
emission of gravitational waves; due to the energy loss 
the orbit contracts, the orbital velocity increases 
and  the system emits more gravitational energy. 
The process of inspiralling proceeds faster and faster
as the system shrinks and finally the two bodies coalesce and merge.
In this section we shall consider only systems
that are far from coalescence.
Since the two stars are far apart, their internal structure
is not affected by the tidal interaction, and it is possible to
treat them as two pointlike masses
revolving around their common center of mass.
In this case  the time varying quadrupole moment is
entirely due to the orbital motion of the two masses, and
the emitted waveform is easily  calculated
by the quadrupole formula (\ref{quad}).

In 1975, Hulse and Taylor applied the quadrupole formalism
to predict the slowing down of the period of  the binary pulsar
PSR 1913+16 \cite{hulsetaylor}.
They found  \op \frac{dP}{dt} = -2.4\cdot 10^{-12},\cl
 in excellent agreement with  the observed value,
\op \frac{dP}{dt} =  -(2.3\pm 0.22)\cdot 10^{-12},\cl
thus providing the first indirect evidence
of the existence of gravitational waves.
But now the question is: can these waves be detected directly?
The quadrupole formula shows that when
the orbit is  circular the radiation is  emitted at twice 
the keplerian orbital frequency
\be
\label{kepfreq}
\nu_k=\frac{1}{P}=\frac{1}{2\pi}
\left(\frac{G M}{R^3}\right)^{1/2},
\ee
where \op M=M_1+M_2\cl is the total mass of the system;
if it is eccentric, as in the case of PSR 1913+16,
\footnote{PSR 1913+16 is composed of
two very compact stars, with masses \op M_1=1.4411~\msun\cl
and \op M_2=1.3874~\msun,\cl revolving around  their center of mass
with an eccentric orbit ($e=0.617139$),  
and keplerian frequency $\nu_k=(\omega_k/2\pi)=3.583\cdot10^{-5}$ Hz.
The system is at a distance  \op D=5\cl kpc from Earth.}
waves will be emitted at frequencies multiple of \op
\nu_k\cl (in this case,  $R$ must be replaced in Eq.~(\ref{kepfreq})
by the semi-major axis),
and the number of  equally spaced spectral 
lines will increase with the eccentricity \cite{pianeti}. 
It is convenient to introduce a characteristic amplitude
which can be compared to the detectors sensitivity
as follows\cite{Kip}
\be
\label{hc}
h_c(n \nu_k,r)=\sqrt{\frac{2}{3}}\left[
\langle \tilde{h}^{(n)}_{\theta \theta}\rangle^2+
\langle \tilde{h}^{(n)}_{\theta \phi}\rangle^2\right]^{1/2},
\ee
where $\langle \tilde{h}^{(n)}_{\theta \theta}\rangle^2$ and
$\langle \tilde{h}^{(n)}_{\theta \phi}\rangle^2$,
are the square of the $n-$th Fourier component of the two 
independent polarizations, averaged over the solid angle
$$\langle \tilde{h}^{(n)}_{\theta \theta}\rangle^2
=\frac{1}{4\pi}
\int{d\Omega~\left|h_{\theta \theta}(n\nu_k,r,\theta,\phi)
\right|^2},\qquad
\langle \tilde{h}^{(n)}_{\theta \phi}\rangle^2=\frac{1}{4\pi}
\int{d\Omega~\left|h_{\theta \phi}(n\nu_k,r,\theta,\phi)\right|^2},$$
and the factor $\sqrt{2/3}$ takes into account the average over orientation.
In the previous expressions, the waveforms have been
expanded in Fourier series\\ $ h(t)=\sum_{n=-\infty}^{n=+\infty} 
h(n\omega_k)e^{-2in\pi t/P},\quad\hbox{and} \quad
h(n\omega_k)=\frac{1}{P}\int_0^P h(t)e^{+2in\pi t/P}dt,$\
where $\omega_k=2\pi\nu_k.$

\begin{figure}
\centerline{\mbox{
\psfig{figure=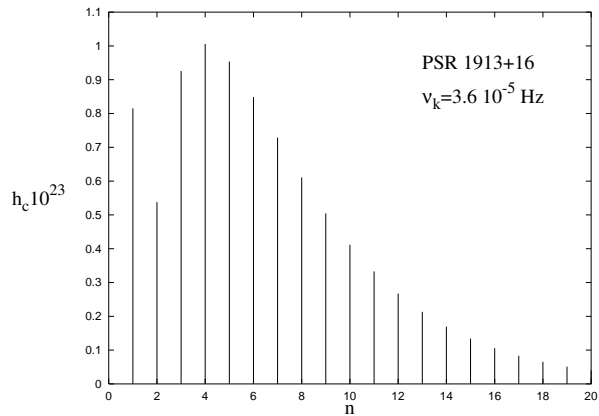,angle=270,width=8cm}
}}
\vskip 10pt
\caption{
The characteristic amplitude of the wave emitted by PSR 1913+16
is plotted versus the harmonic index $n.$
The spectral lines are emitted  at frequencies
multiple of $\nu_k.$
}
\label{fig1}
\end{figure}

The characteristic amplitude \op h_c\cl  emitted by
PSR 1913+16 is shown in  figure 1
as a function of the harmonic index $n$.
The maximum occurs at
$ \nu_{max}=1.44\cdot 10^{-4}~Hz,$ 
and the amplitude of the corresponding line is
$h_{c~max}\sim  10^{-23}.$

Thus the emission of this system is in the bandwidth of 
the space-based interferometer LISA, but 
the expected sensitivity curve  of this instrument
shows that the signal is too low to be detectable even
with five years of integration.

However,  other interesting sources of radiation
of this kind exist in our Galaxy.  
For instance, in a recent paper the gravitational emission of
cataclysmic variables has been considered \cite{catvariab}. 
These systems are semi-detached binaries of low mass and very short period, in
which the primary star is an accreting degenerate white dwarf,
and the secondary is usually a late type-star filling its Roche
lobe and transfering matter on the companion.
In table I the masses, emission frequencies and characteristic wave amplitude
are listed for some of these short period systems; among the systems analized in 
ref. 11,  we have selected only  those 
that emit gravitational waves large enough to  
be detectable by LISA with one year of integration.

\begin{table}[t]
\caption{Cataclysmic variables with component masses $m_1$ and
$m_2$, gravitational  wave frequency $\nu_{GW}$,
and gravitational wave amplitude $h_c$. }
\vspace{0.2cm}
\begin{center}
\label{table2}
\begin{tabular}{@{}lccccccccccc@{}}
\multicolumn{1}{c}{name}&&$m_1(M_\odot)$ &&$m_2 (M_\odot)$&  
& $\nu_{GW}$ (Hz) &&  $\log h_c$\\
\hline  
U Gem  &&1.26  &&   0.57&&$1.31\cdot 10^{-4}$&& -20.8 \\
IP Peg &&1.15  &&   0.67&&$1.46\cdot 10^{-4}$&& -20.9 \\
HU Aqr &&0.95  &&   0.15&&$2.67\cdot 10^{-4}$&& -21.3 \\
VW Hyi &&0.63  &&   0.11&&$3.12\cdot 10^{-4}$&& -21.3 \\
EX Hya &&0.78  &&   0.13&&$3.39\cdot 10^{-4}$&& -21.4 \\
WZ Sge &&0.45  &&   0.058&&$4.08\cdot 10^{-4}$&& -22.1 \\
ST LMi &&0.76  &&   0.17&& $2.93\cdot 10^{-4}$&& -21.4 \\
SW UMa &&0.71  &&   0.10&& $4.93\cdot 10^{-4}$&& -21.6 \\
Z Cha &&0.84   &&   0.125&& $3.11\cdot 10^{-4}$&& -21.5 \\
V 436 Cen &&0.7  &&  0.17&& $3.70\cdot 10^{-4}$&& -21.6 \\
\hline\hline
\end{tabular}
\end{center}
\end{table}

\section{Extrasolar planetary systems}\label{sec4}
In recent years a large number of extrasolar planetary systems 
have  been discovered  at  very short distance from us
\cite{webplanets}, which exhibit  some very interesting properties.
They are very close to Earth (less than 10-20 pc), and 
they are formed by a solar type star  and  one or more orbiting  companions, which can be
planets, super-planets or, in some cases, brown dwarfs.
Many of these companions are orbiting the main star  at such short distance that conflicts
with the predictions of the standard theories of planetary 
formation and evolution.  For instance, some planets 
\footnote{Here and in the following we shall indicate as  ``planet"
also brown dwarfs and super planets.}
have orbital periods shorter than three days
(for comparison, Mercury's period is 88 days).
In the light of these findings, it is interesting to ask the 
following question:
is it possible that a planet orbit the main star so close as to excite one of its 
modes of oscillations?
How much energy would be emitted in the form of gravitational 
waves by a system in this resonant condition, compared  to the 
energy emitted because of the
orbital motion? For how long would this condition persist?
These questions have been analysed in two recent paper 
\cite{pianeti} \cite{gmodes},
whose results I shall briefly summarize. 
First of all we computed the frequencies of pulsation of the modes 
of a solar type star,
by integrating the equation of stellar perturbations in the framework of General
Relativity. We considered a polytropic model of star, $p=K\epsilon^{1+1/n}$ with 
$n=3,$ adiabatic  exponent $\gamma=5/3,$ central density \op\epsilon_0=76~g/cm^3,\cl 
and $c^2\epsilon_0/p_0=5.53\cdot 10^5.$ This model  gives a star with the
same mass and radius as the Sun \cite{gmodes}.  
In table 2 the values of the frequencies of the 
fundamental mode and of the first g-modes of the star are tabulated for
$\ell=2.$ 
We do not include the frequencies of the p-modes because they
are irrelevant to the following discussion.

\begin{table}[t]
\centering
\caption{
Values of the frequencies (in $\mu$Hz) of the fundamental and of the first 
g-modes of oscillation of a solar-mass polytropic star with $n=3,$ for $\ell=2$.
}
\begin{tabular}{@{}llllllllllll@{}}
\hline
                         &\multicolumn{8}{c}{} \\
			 & f-mode
			 & $g_{1}$ & $g_{2}$ & $g_{3}$ & $g_{4}$ & $g_{5}$
			 & $g_{6}$ & $g_{7}$ & $g_{8}$ & $g_{9}$ & $g_{10}$  \\
			 \hline
	 $\nu_i$ &$285$ &$221$ &$168 $&$135$&$113$&$97 $
			 &$85$ &$75$  &$68 $ &$62 $&$57 $ \\
			 \hline
			 \hline
			 \end{tabular}
			 \label{freqs}
			 \end{table}
If a planet moves on an orbit of radius $R$
(we shall assume for simplicity that the orbit is
circular), the keplerian orbital frequency is
\op
\label{kepler}
\nu_k=\frac{1}{2\pi}\sqrt{\frac{G(M_\star+M_p)}{R^3}},
\cl
where $M_\star$  and $M_p$ are the mass of the star and of 
the planet, respectively.
The planet can excite a mode of frequency $\nu_i,$ only if  $\nu_i$ is 
a specific multiple of the orbital frequency, i.e.
\be
\label{reso}
\nu_i =\ell~\nu_k,
\ee
where $\ell$ is the considered multipole.
Introducing the dimensionless frequency
\[
k_i=  \nu_i/ \sqrt{\frac{G M_\star}{R_\star^3}};
\]
where $R_\star$ is the radius of the star, 
the resonant condition (\ref{reso}) 
can be written as
\be
R_i=\left[
\frac{\ell^2}{k^2_{i}}\cdot \left(1+\frac{M_p}{M_\star}\right) \right]^{1/3}R_\star.
\label{cond2}
\ee
Thus, $R_i$ is the value of the  orbital radius
for a given mode to be excited; for instance, using Eq.
(\ref{cond2}) we find that  in order
to excite the fundamental mode of the considered star,
the planet should move on an  orbit of  radius
smaller than 2 stellar radii, therefore we first
need to verify whether a planet can move on such close orbit without being
disrupted by the tidal interaction.
This is equivalent to establish at which distance  the star 
starts to accrete matter 
from the planet (or viceversa),   i.e. when
the planet or the star overflow their Roche lobes
\footnote{Following the analysis in ref. \cite{melting},
temperature effects, which may provoque the melting of the 
planet or the evaporation of its atmosphere, can be shown 
to be less stringent than the Roche-lobe limit.}. 

It is an easy exercise to compute the radius of the
Roche lobe, $R_{RL}$,
in newtonian theory, which fully applies to the case we are examining.
If we choose a reference frame co-rotating with the two masses, with origin on the center
of mass  and such that the x-y plane coincides with the orbital plane, and the 
x-axis joins the two masses,
the newtonian potential of the star-planet system can be written as
\[
U(x,y)=-\frac{GM_p}{{\bf x}-{\bf x_p}}-
\frac{GM_\star}{{\bf x}-{\bf x_\star}}
-\frac{1}{2}\omega_k^2 \vert{\bf x}\vert^2,
\]
where $x_\star=\left(\frac{M_p}{M_p+M_\star}R,0\right),$
$x_p=\left(-\frac{M_\star}{M_p+M_\star}R,0\right),$ 
$R$ is  the orbital radius of the planet, and the last term is
the centrifugal potential.
A gas particle at a large radius on the planet
may feel a gravitational force from the star
that is comparable to, or even larger, than that from the
planet. In that case,
the gas particle is unstable against being transferred from the
planet to the star.
This can be understood by plotting contours of equal gravitational
potential: if the potential  is large and negative,
the equipotential surfaces are two
spheroids, one centered on $M_p$, the other on $M_\star.$
As the potential increases the spheroids  deform, and finally
touch in one point lying on the axis which joins
the star and the planet; the surface they form in this
configuration is the Roche lobe surface,  the unique point
where the  potential energy contour
intersects itself is  called the inner Lagrange
point, and the  curve which encloses  the Roche lobes is said ``first
Lagrangian curve".
If the planet expands to fill its Roche lobe,
the matter near the inner Lagrange point will spill  over
onto the star.
Similarly, if the star fills its own 
Roche lobe it can accrete matter onto the planet.
We can define the Roche lobe radius for the planet, $R_{RL},$ 
as the radius of the circle centered on the planet  and tangent to the first
Lagrangian curve (similarly for the star). 
$R_{RL}$ can be found numerically, and it depends exclusively on the planet orbital
radius $R$ and on the star and planet masses.

The procedure to establish if a planet is allowed to orbit at a given distance 
from the star exciting its modes is the following:\\
- we assign a value of $R=R_i$ which corresponds to 
the excitation of a given mode, according to  Eq. (\ref{cond2});\\ 
- we determine the Roche lobe radius
$R_{RL}(R_i,M_p,M_\star)$ and compute  the dimensionless quantity
\op \overline{R}_{RL}\cl  given by
\be
\overline{R}_{RL}=R_{RL}/R_i;
\label{rbar}
\ee
- if the planet radius would be equal to $R_{RL},$ its
average density would have a critical value, $\rho_{RL},$ given by
\be
\label{critdens}
\rho_{RL} =
\frac{M_p}{\frac{4}{3}\pi R_{RL}^3}.
\ee
Thus, if $\rho_p >  \rho_{RL}$ the planet will not fill its Roche lobe and will not
accrete matter on the star.  Using Eqs.
 Eqs.~(\ref{critdens}),(\ref{cond2}) and (\ref{rbar}),
the value of the critical density can be rewritten as
\be
\rho_{RL}=\frac{M_p}{
\frac{4}{3}\pi R_\star^3
\left[\dps{\frac{\ell^2}{k^2_{i}}\cdot
\left(1+\frac{M_p}{M_\star}\right)}
\right]\overline{R}_{RL}^3},
\ee
or, normalizing to the mass of the central star, $\rho_\star,$
\be
\frac{\rho_{RL}}{\rho_\star}
=k^2_{i}\cdot \frac{M_p/M_\star}{\ell^2\left(1+M_p/M_\star\right)
\overline{R}_{RL}^3}.
\label{cond4}
\ee
In conclusion, a planet can  excite the ith-mode of the star without overflowing
its Roche lobe, only if the ratio between its mean density  and that
of the central star  is larger than the  critical ratio (\ref{cond4}).

 \begin{table}
 \centering
 \caption{
The critical ratio $\frac{\rho_{RL}}{\rho_\star}$ is given for  three planets with mass
equal to that of the Earth ($M_E$), of Jupiter ($M_J$) and
of a brown dwarf with $M_{BD}=40~M_J$, and for the different modes.
This critical ratio  corresponds to the 
minimum mean density that a planet should have in order to be 
allowed to move on an orbit which
corresponds to the excitation of a g-mode,
without being disrupted by tidal interactions.
 }
 \vskip 10pt
 \begin{tabular}{@{}lllllllllllll@{}}
&\multicolumn{11}{c}{$\frac{\rho_{RL}}{\rho_\star}$} \\
\hline
&
& $g_{1}$ & $g_{2}$ & $g_{3}$ & $g_{4}$ & $g_{5}$
& $g_{6}$ & $g_{7}$ & $g_{8}$ & $g_{9}$ & $g_{10}$ \\
\hline&$M_{E}$
&12.5 &7.21 &4.64 & 3.24 & 2.38
& 1.83 & 1.45 & 1.17 & 0.97 & 0.82 \\
\hline
&$M_{J}$
&13.0 &7.49 &4.82 &3.37 &2.48
&1.90 &1.51 &1.22 &1.01 & 0.85 \\
\hline
&$M_{BD}$
&-    &-    &-    & 3.68 & 2.71
& 2.08 & 1.65 & 1.34 & 1.11 & 0.93 \\
\hline
\hline
\end{tabular}
\label{lobi}
\end{table}
   
We have computed the critical ratio (\ref{cond4})
assuming that the central star has
three different companion: two planets with
the mass  of the Earth and of Jupiter,  and a
brown dwarf  of 40 jovian masses,
$M_E,$ $M_J$ and $M_{BD},$ respectively, imposing that they are on an orbit which
corresponds to the resonant excitation of
a given mode of the solar type star.
We do not consider smaller planets because they produce gravitational signals that are 
too small to be interesting.
The results of this analysis are shown in Table  \ref{lobi},
where  we tabulate the value of $\frac{\rho_{RL}}{\rho_\star}$
for the three considered  companions.

A planet like the Earth has a mean density such that $\rho_{Earth}/\rho_\odot=3.9,$
whereas  for a  planet like Jupiter $\rho_{J}/\rho_\odot=0.9;$
thus, from Table \ref{lobi} we see that
an  Earth-like planet can orbit sufficiently close to the 
star to excite g-modes of order higher or equal to $n=4$, 
whereas a  Jupiter-like planet
can excite only the mode $g_{10}$ or higher.
We find that in no case the fundamental mode can be excited 
without the planet being disrupted by accretion.
According to  the brown dwarf model (``model G'') by Burrows
and Liebert \cite{burrows} an evolved, 40 $M_J$
brown dwarf has a radius $R_{BD}=5.9\cdot 10^4$ km, and a 
corresponding mean density $\rho_{BD}=88$ g$\cdot$cm$^{-3}$; 
consequently
$\rho_{BD}/\rho_\odot=64$, and this value is high enough to
allow a brown dwarf companion to excite all the g-modes of the central
star.  However,  we also need to take into account
the destabilizing mechanism  of  mass
accretion {\it from} the central star. This imposes a further constraint,
and this is the reason why the slots corresponding to the
excitation of the g-modes lower  that g$_4$ in the last row of Table
\ref{lobi} are empty.

Having established that some g-modes of the central star can, in principle, be
excited, we turn to the next question, i.e. :
how much energy would be emitted in  gravitational waves by a system
in this resonant condition, compared  to the energy emitted because of the
orbital motion? 
The amplitude of the wave emitted because of the orbital 
motion can be computed by using the quadrupole formula (\ref{quad}); 
in the case of circular orbit the
characteristic amplitude (\ref{hc})  becomes
\be
\label{hq}
h_Q(R_i)=
\frac{4M_p}{r}~\sqrt{\frac{2}{15}}
\left(\omega_k R_i\right)^2.
\ee
The values of this amplitude
for the three planets considered above,  are shown in table \ref{amplitudes}. 
\begin{table}
\centering
\caption{ The amplitude of the gravitational signal emitted
when  the three companions considered in Table \ref{lobi}
move on a circular orbit
of radius $R_i$, such that the condition of resonant excitation of a
g-mode  is satisfied,
is computed by the quadrupole formalism (see Eq. (\ref{hq}))
for the modes allowed  by the
Roche-lobe analysis.
The planetary systems are assumed to be at a distance of
$10$ pc from Earth.
}
\begin{tabular}{@{}lllllllll@{}}
			 \hline
			 &&  $g_{4}$ & $g_{5}$
			 & $g_{6}$ & $g_{7}$ & $g_{8}$ & $g_{9}$ & $g_{10}$ \\
			 \hline
			 \\

$h^{Earth}_Q\cdot 10^{26}$  &
			 
			 & $3.0$ & $2.7$ & $2.5$ & $2.3$ & $2.2$ & $2.0$ & $1.9$  \\
			 \hline
			 \\
$h^{Brown Dwarf}_Q\cdot 10^{22}$& 			 
			 & $3.9$ & $3.5$ & $3.2$ & $3.0$ & $2.8$ & $2.6$ & $2.4$  \\
			 \hline
			 \\
$h^{Jupiter}_Q \cdot 10^{24}$& 
			 & - &  - &  - &  - &  - &  - & $6.1$\\
			 \hline
			 \hline
			 \end{tabular}\label{amplitudes}
			 \end{table}
However, if the stellar modes are excited the star will pulsate and emit more
gravitational energy than that which can be computed by the quadrupole formula.
In order to evaluate the resonant contribution we need to follow a different
approach which takes into account the internal structure of the main star.
One possibility is to use a perturbative approach which consists in the following.
The main star is assumed to be an extended body, whose equilibrium structure is described by
an exact solution of the relativistic equations of  hydrostatic equilibrium;
the planet is considered as a pointlike mass which  induces
a perturbation on the gravitational field and on the thermodynamical structure 
of the star.
Under these assumptions  we solve the equations of stellar perturbations to compute
the characteristic  amplitude
of the gravitational wave emitted when the planet
moves  on  a circular orbit close to a resonance \cite{gmodes}. In the following, we shall
call this amplitude $h_R,$ to indicate that it has been computed by solving the
relativistic equations of stellar perturbations.

\begin{figure}
\centerline{\mbox{
\psfig{figure=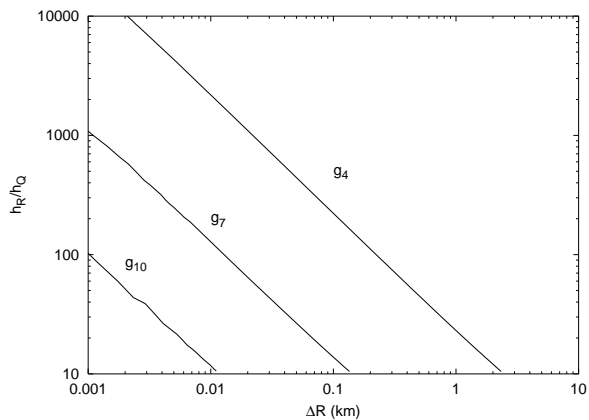,angle=270,width=8cm}
}}
\vskip 10pt
\caption{
The logarithm of the ratio $h_R^{l=2}(R_i+\Delta R)/h_Q(R_i)$
is plotted as a function of $ \Delta R$ for the modes g$_4$,
g$_7$, and g$_{10}$.
}
\label{gmodesexc}
\end{figure}

We have computed  $h_R$ for $\ell=2$ (which is the relevant contribution),
assuming that the planet moves on a circular orbit of radius
$R_0=R_i+\Delta R,$  where $R_i$ is the radius corresponding to
the resonant excitation of the  mode $g_i$, for the modes allowed by the Roche
lobe analysis.  We find that, as the planet approaches the resonant orbit,
$h_R$ grows very sharply.
In figure  \ref{gmodesexc} we plot the ratio $h_R(R_0)/h_Q(R_i)$
as a function of \op \Delta R,\cl to see  how much the amplitude of the
emitted wave increases, due to the excitation of a g-mode,
with respect to the quadrupole emission.
The plot is done  for the modes g$_4$,
g$_7$, and g$_{10},$ and it shows a power-law behaviour nearly independent of the
order of the mode.
It should be stressed that $h_R(R_0)/h_Q(R_i)$
is independent of the mass of the planet, but depends on the selected stellar model.
From figure \ref{gmodesexc} we see that as the planet approaches
a resonant orbit  the amplitude of the emitted wave may
become significantly higher than that computed by the quadrupole formalism,
 $h_Q$. Thus the next question to answer is how long
can a planet  orbit near  a resonance,
before radiation reaction effects move it off.
Indeed, the loss of energy in gravitational waves causes a shrinking
of the orbit of a planetary system, and the efficiency of this process
increases as the planet approaches a resonant orbit.
We shall now compute the time a planet takes to move from an orbit of radius
$R_0=R_i+\Delta R,$  where  the amplification  factor
$h_R/h_Q$ has some  assigned value,
to the resonant orbit $R_i$, because of
radiation reaction effects.
This timescale will indicate whether a planet can stay in the
resonant region long enough to be possibly observed.
On the assumption that the timescale over which the orbital radius evolves
is much longer than the orbital period (adiabatic approximation),
the  orbital shrinking  can be computed  from  the energy conservation law
\be\label{energycons}
M_p\left\langle\frac{dE}{dt}\right\rangle
+\left\langle\frac{dE_{GW}}{dt}\right\rangle=0,
\ee
where $E$ is the energy per unit mass of the planet which moves on the geodesic of radius
$R_0$
\be
\label{energyandmom}
E=\left(1-\frac{2 M_\star}{R_0}\right)\left(1-\frac{3 M_\star}{R_0}\right)^{-1/2},
\ee
and $\left\langle\frac{dE_{GW}}{dt}\right\rangle$ is
the energy emitted in gravitational waves,
computed  numerically by using the perturbative approach (see ref. \cite{gmodes} for
details).
Since
$\left\langle\frac{dE}{dt}\right\rangle=
\left\langle\frac{dR_0}{dt}\right\rangle/\left\langle\frac{dR_0}{dE}\right\rangle$,
using  Eq. (\ref{energyandmom}) 
Eq. (\ref{energycons}) gives
\be
\left\langle\frac{dR_0}{dt}\right\rangle=-\frac{2 R_0^2}{M_p M_\star}
\frac{(1-3M_\star/R_0)^{3/2}}{(1-6M_\star/R_0)}
\left\langle\frac{dE_{GW}}{dt}\right\rangle,
\ee
from which the time needed  for the planet to reach the resonant orbit
can be computed
\be
\label{deltat}\Delta T=
-\frac{M_p M_\star}{2}
\int_{R_i+\Delta R}^{R_i}
\frac{(1-6M_\star/R_0)}{(1-3M_\star/R_0)^{3/2}}
\frac{dR_0}{R_0^2~\left\langle\frac{dE_{GW}}{dt}\right\rangle}.
\ee
It should be noted that since $\left\langle\frac{dE_{GW}}{dt}\right\rangle$
is proportional to $M_p^2$, \op \Delta T\cl is longer for  smaller
planets.
We have computed $\Delta T$ for the three companions $M_E,$ $M_J$ and
$M_{BD},$
\footnote{
The data in table 5 have been obtained by repeating the calculations described in ref.
\cite{gmodes}, after correcting an error found in the numerical code. The qualitative
results of ref. \cite{gmodes}  are, however, correct.
}
by the following steps:\\
- we assign a value of the orbital radius $R=R_i$ 
which corresponds  to the excitation of a g-mode allowed 
by the Roche lobe analysis\\
- we compute the radius of the orbit $R_0=R_i+\Delta R$ for which
the amplification
factor \op A= h_R(R_0)/h_Q(R_i)\cl has an assigned value\\
- we  compute the energy radiated in gravitational waves,
$\left\langle\frac{dE_{GW}}{dt}\right\rangle,$ as a function of the orbital radius,
in the range between $R_0$ and $R_i$\\
- using Eq. ~(\ref{deltat}) we compute $\Delta T$, which tells us how long can the
planet orbit the star in the resonant region between $R_0$ and
$R_i$, emitting a wave of
amplitude higher than $[A\times h_Q(R_i)]$.\\
The results are summarized in table \ref{times}.
These data have to be used together with  those  in table
\ref{amplitudes} as follows.
Consider for instance a planet like the Earth, orbiting its
sun on an orbit resonant with the mode g$_4$.
According to the quadrupole formalism
it would emit a signal of amplitude $h_Q(R_4)=3\cdot
10^{-26},$\ (table \ref{amplitudes}, first row)
at a frequency $\nu_{GW}=1.13\cdot 10^{-4}$ Hz
(table \ref{times}, first row) .
The data of table \ref{times}, which include the resonant contribution
to the emitted radiation,
indicate  that before reaching that resonant orbit of radius $R_4$,
the Earth-like planet would orbit in a region  of thickness
$\Delta R =1.7$ km
slowly spiralling in, emitting waves with amplitude
$h_R > 10 h_Q(R_4)=3\cdot 10^{-25},$\ for a time interval
of $5.4\cdot 10^6$ years, and that while spanning the smaller radial
region  $\Delta R =312$ m,
the emitted wave  would reach an amplitude
$h_R > 50 h_Q(R_4)=1.5\cdot 10^{-24},$\ for a time interval
of $3.8\cdot 10^4$ years.
\begin{table}[t]
\caption{
For each mode allowed by the Roche lobe analysis (see text), we give
the frequency  of the wave
emitted when a planet  moves  on an orbit resonant with
a g-mode  of its sun (column 2). 
When the planet  spans a radial region of thickness $\Delta R$ (column 3)
approaching a resonant
orbit, due to the stellar pulsations the amplitude of the emitted wave is amplified,
with respect to the quadrupole amplitude, by
a factor greater than $A$ (column 4).
In the last three columns we give  the time interval  $\Delta T$ needed for
the three companions to  span the region $\Delta R$ and
reach the resonance.
}
\vspace{0.2cm}
\begin{center}
\label{table3}
\begin{tabular}{@{}cllllll@{}}
\\
\multicolumn{1}{c}
{Mode}&$\nu_{GW}$ ($\mu$Hz)& $\Delta R$(m)&
$A$&$\Delta T_E$(yrs)&$\Delta T_{BD}$(yrs)&$\Delta T_{J}$(yrs)\\
\hline
g$_4$ &$113$      &1700  &$10$   &$5.4\cdot10^6$ &$4.3\cdot10^2$         &-\\
&                 &312  &$50$   &$3.8\cdot10^4$ &$3.0$                  &-\\
\hline
g$_5$ &$97$       &616    &$10$ &$2.7\cdot10^6$ &$2.1\cdot10^2$         &-\\
&                 &113  &$50$    &$1.9\cdot10^4$ &$1.5$                  &-\\
\hline
g$_6$ &$85$       &240    &$10$ &$1.4\cdot10^6$ &$1.1\cdot10^2$         &-\\
&                 &44  &$50$    &$9.6\cdot10^3$ &$7.5\cdot10^{-1}$      &-\\
\hline
g$_7$  &$75$      &98      &$10$&$7.0\cdot10^5$ &$55$ &-\\
&                 &18  &$50$     &$5.0\cdot10^3$ &$3.9\cdot10^{-1}$      &-\\
\hline
g$_8$ &$68$       &42    &$10$  &$3.8\cdot10^5$ &$30$ &-\\
&                 &8  &$50$     &$2.5\cdot10^3$ &$1.9\cdot10^{-1}$ &-\\
\hline
g$_9$ &$62$       &18      &$10$ &$1.9\cdot10^5$ &$15$ &-\\
&                 &3  &$50$     &$1.5\cdot10^3$ &$1.2\cdot10^{-1}$ &-\\
\hline
g$_{10}$ &$57$    &8     &$10$  &$1.0\cdot10^5$ &$7.8$    &$3.1\cdot 10^2$ \\
&                 &1  &$50$      &$6.7\cdot10^2$ &$5.3\cdot10^{-2}$     &$2.1$\\
\hline\hline
\end{tabular}
\end{center}
\label{times} 
\end{table}
A jovian planet, on the other hand, could only excite modes of order
$n=10$ or higher, which would correspond to a resonant frequency of
$\nu_{GW}=5.7\cdot 10^{-5}$ Hz and a
gravitational wave amplitude greater than \ $6\cdot 10^{-23}$ for  310
years ($\Delta R =8$ m), or greater than \ $3\cdot 10^{-22}$ for $\sim 2$ years ($\Delta R
=1$m).
From these data we see that the higher  the order of the mode, the more
difficult it is to excite it, because the region where the resonant
effects become significant gets narrower and the planet transits through
it for a shorter time.
Much more interesting are the data for a brown dwarf companion.
In this case, for instance, the region that would correspond
to the resonant excitation of the mode g$_4$ ($\Delta R=1.7$ km),
with a wave amplitude greater than $3.9\cdot 10^{-21}$, would  be
spanned in $430$ years,
whereas the emitted wave would have an amplitude greater than
$\sim 2\cdot 10^{-20}$ ($\Delta R=312$ m)
over a time interval of $\sim 3$ years.
The results of this study show that if  a brown dwarf would orbit a solar type star
at such distance that a g-mode is excited, the emitted 
radiation may be strong enough, and for a sufficiently long time interval,
to be detectable by LISA.

\section{Coalescing binary systems}\label{sec5}

The main target of ground-based interferometers is to detect the signal emitted
during the late phases of inspiraling of  a binary system.
Assuming that the two coalescing bodies are two point masses, $M_1$ and $M_2$,
by using the quadrupole formalism and including the radiation reaction effects (along the
lines discussed  in Sec. 4), it is possible to show that the loss of 
gravitational energy induces a circularization of the
orbit \cite{PetersMathews}, and that the radius decreases according to the law
\op
R(t)=R_{in}\left(1-t/t_{coal}\right)^{1/4};
\cl
where \op t_{coal}=\frac{5}{256}\frac{R_{in}^4}{\mu M^2}\cl is the time of the
final coalescence, and 
\op\mu=M_1 M_2/M\cl is the reduced mass of the system of total mass $M=M_1+M_2$.
The orbital frequency consequently increases, and so does the
frequency of the emitted wave,
with a time dependence given by
\op
\nu=\frac{1}{\pi}\left[
\frac{5}{256}\frac{1}{\mu M^{2/3}}\frac{1}{(t_{coal}-t)}
\right]^{3/8}.
\cl
Thus, the gravitational signal emitted by a coalescing system resembles 
the chirp of a singing bird.

At some point of the evolution, for instance $\sim 10^8$ years after
formation if the  binary system is composed of two neutron stars,
the frequency of the emitted signal enters  the
bandwidth of the ground based interferometers, and in about 15-20 minutes
(depending on the mass of the stars)
sweeps  the frequency region ranging from $\sim10$Hz
up to $\sim900$Hz.
The amplitudes of the two polarizations of the wave emitted 
during this fast inspiralling phase are
\be
\label{wavetrain}
h_+= 
\frac{2(1+\cos^2 i)
\mu\left(\pi M\nu\right)^{2/3}}{r}
\cos(2\pi \nu t),\quad
h_\times=\pm \frac{4\mu\cos i \left(\pi M\nu\right)^{2/3}}{r}
\sin(2\pi \nu t),
\ee
where $i$ is the angle of inclination of the orbit to the line of sight,
and the emitted energy per unit frequency
is given by
\op
\frac{dE}{d\nu}=\frac{\pi^{2/3}}{3}\mu\frac{M^{2/3}}{\nu^{1/3}}.
\cl

This description of the coalescence based on the quadrupole formalism
breaks down when the two stars are so close that tidal effects become important.
For instance, for binaries with one white dwarf companion,
the maximum frequency, i.e. the minimum distance
between the two stellar components, is set in order to cut-off the
Roche-lobe contact stage. In fact, the mass transfer from one component
to its companion transforms the original detached binary into a
semi-detached binary. This process can be accompanied by loss 
of angular momentum with mass loss from the system and the above 
description cannot be applied.
In this case,  the mimimum orbital separation that two white dwarfs can reach 
is given by $r_{wd}(M_1)+r_{wd}(M_2),$
where
\op
r_{wd}(M)=0.012 R_{\odot} \sqrt{\left(M/1.44
M_{\odot}\right)^{-2/3} -
\,\,\left(M/1.44 M_{\odot}\right)^{2/3}}
\cl
is the approximate formula for the  radius of a white dwarf
\cite{Nauenberg}.
If the system is composed, say,  by two white dwarfs of $0.9~M_\odot$,
the cutoff frequency is $\nu_{GW~max}\sim 0.1$ Hz. These signals are
therefore candidates for detection by spaceborne interferometers. 

The situation is different for NS-binaries or BH-binaries because, 
being the two bodies extremely compact, the
pointlike approximation described above works remarkably well up to much
higher frequencies,  where the post-newtonian formalism can be applied
to correct the waveforms emitted during the latest phases of the inspiral;
however, even this  more refined waveforms fail to
correctly reproduce the emitted signal when the innermost stable circular
orbit (ISCO) is approached, particularly when the coalescing bodies are neutron stars.
Indeed, in a recent paper we have shown that  due to a resonant excitation of
the modes of oscillation of the stars, the emitted energy may be enhanced with respect 
to the orbital contribution \cite{tutti2}.
This phenomenon emerges  only if we use an approach in which
the stars are allowed to have an internal structure, which plays a dynamical  
role by exchanging energy with the gravitational field.
For instance, in \cite{tutti2}
we have used the  same perturbative approach described in Sec. \ref{sec4} to
study the emission of extrasolar planetary systems: we assume that
one of the two neutron  is a ``true" neutron star,
whose equilibrium structure is governed by the TOV equations 
of  hydrostatic equilibrium, whereas the second star is a point mass which  perturbes
the gravitational field and the fluid of the extended companion. 
To extract the effects that the
internal structure of the star has on the gravitational emission 
we have selected a set of parameters which allow to  encompass
a reasonable range of stellar models, and to cover most of the range of structural
properties obtained with realistic EOS.
The values of the parameters are given in Table \ref{table6}.
\begin{table}[t]
\caption{
Parameters of the polytropic stars we consider in our analysis:
the polytropic index $n$, the central density, the ratio
$\alpha_0=c^2 \epsilon_0/p_0$ of central 
density to central pressure, the mass,  the radius  and the ratio $R/M$
(in geometric units).
The central density is chosen in such a way that the stellar mass is equal to
$1.4 M_\odot$, except for model A, the mass of which is about one solar mass.
}
\vspace{0.2cm}
\begin{center}
\begin{tabular}{@{}clllllr@{}}
\hline
Model number &$n$ &$\epsilon_0$ (g/cm$^3$) &$\alpha_0$  &$M$ ($M_\odot$) &$R$ (km)&
$R/M$\\
\hline
A   &1.5     & $1.00\times 10^{15}$    &13.552     &$0.945$     &14.07 & 10.08\\
B   &1       & $6.584\times 10^{14}$   &9.669      &$1.4$       &15.00 & 7.26\\
C   &1.5     & $1.260\times 10^{15}$   &8.205      &$1.4$       &15.00 & 7.26\\
D   &1       & $2.455\times 10^{15}$   &4.490      &$1.4$       &9.80 & 4.74\\
E   &1.5     & $8.156\times 10^{15}$   &2.146      &$1.4$       &9.00 & 4.35\\
\hline
\hline
\end{tabular}
\end{center}
\label{table6}
\end{table}
For each model, we have integrated the equations of stellar perturbations
in General Relativity, and   we have
 computed the energy flux emitted in gravitational
waves,  \op  \dot{E}^R,\cl
assuming that the perturbing point mass, $m_0,$ is moving on a 
circular orbit of radius \op R_0,\cl with orbital
velocity $v,$ and semilatus rectum $p$, given by
\be
v=\left( M\omega_k\right)^{1/3}=\frac{1}{\sqrt{p}},
\label{velocity}
\ee
where $M$ is the mass of the star.
\begin{figure}
\centerline{\mbox{
\psfig{figure=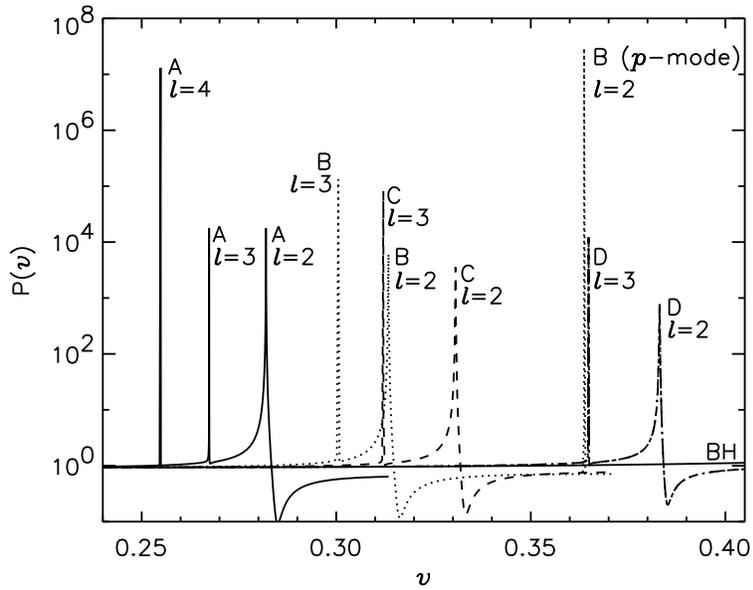,width=11cm},
}}
\vskip 10pt
\caption{
The normalized energy flux, $P(v),$ is plotted as a function of the orbital
velocity for the stellar models  given in Table \ref{table6}
and for a black hole.
For model D and for the black hole the curves extend
up to the velocity which correspond to the ISCO, whereas for the other models
they stop when the mass $m_0$ reaches the surface of the star.
The sharp peaks indicate that, for different values of the harmonic index $l,$
the fundamental quasi-normal modes of the star are excited if the orbital
frequency satisfies the resonant condition (\ref{reso}); the curve of the
stellar model B has a  peak at high $v$ which correspond to the
excitation of the first p-mode for $l=2.$
The most compact model E is not shown in the figure because at this scale
it is indistinguishable from the black hole.
}
\label{fig3}
\end{figure}
The results are shown in  Fig.  \ref{fig3}, where we plot  the normalized
energy flux, \op P \equiv \dot{E}^R/\dot{E}^N,\cl
where $\dot E^N$ is the orbital contribution computed by the
quadrupole formalism
\be
\label{newtflux}
\dot E^N=\frac{32}{5}\frac{m_0^2}{M^2} v^{10}~.
\ee
$P(v)$ has been obtained by adding the contributions of different $l$ and $m$, with
\op 2 \leq l \leq 7,\cl and it is plotted in  Fig.  \ref{fig3}
as a function of the orbital
velocity, for the models of star we have considered;  for comparison, we also
plot $P(v)$ computed in the case when the perturbed object is a  black hole.
In this case the plot extends up to the velocity which corresponds to the 
ISCO, \op R_0=6 M,\op whereas for the stellar models
it  stops when  $m_0$ reaches the surface of the star.
From Fig.  \ref{fig3} we see that sharp peaks appear  if the central object is 
a star: they correspond to the
excitation of the fundamental quasi-normal modes of the star for different values of
the harmonic index $\ell$.
In the case of model B the first p-mode for $\ell=2$ is also excited.
Since the  peaks of the mode excitation are very high,
the scale chosen on the vertical axis of Fig. \ref{fig3} makes
the response of the black hole to appear as a flat line.
The reason is that, since
the frequency of the lowest quasi-normal modes of a black hole are 
much higher than those of a star with the same mass,
the circular orbit that would excite them would have a radius smaller
than $6 M_{BH}.$ Thus, in the range of $v$ considered in Fig. \ref{fig3} the energy flux
emitted by the black hole is due essentially to the orbital motion.
In Fig. \ref{fig4}, we show a zoom of Fig. \ref{fig3}
restricted to the region $v < 0.28,$ which is  far enough from the
resonant orbits (except that for model A). 
In this case
we can appreciate the differences between the emission of different stellar
models and that of a black hole.
If the orbital velocity is smaller than 0.16 all curves are practically
indistinguishable.
Fig. \ref{fig4} shows that the normalized energy fluxes emitted by
different stellar models have a different slope, and are always larger than the
flux emitted by the black hole.
The increase in the energy output at orbital velocities larger than  $v=0.16$ 
is just an effect of the resonance tail.
\begin{figure}
\centerline{\mbox{
\psfig{figure=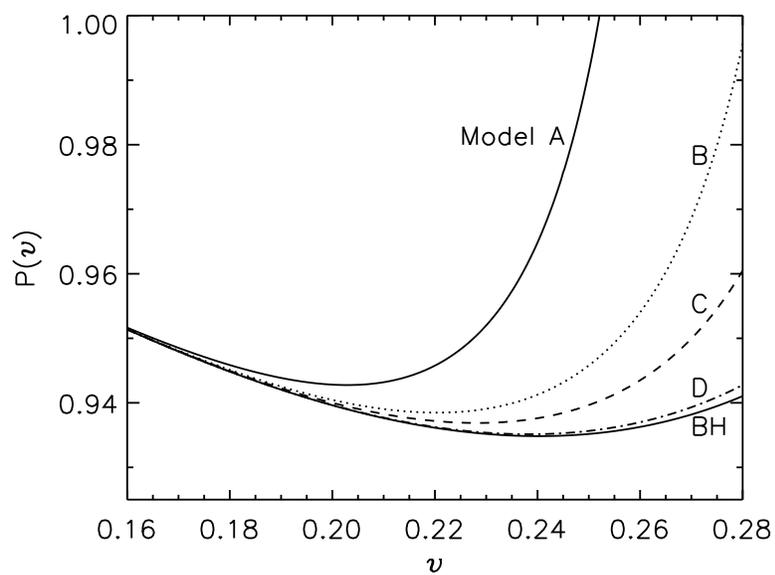,width=11cm},
}}
\vskip 10pt
\caption{
The normalized energy flux, $P(v),$ is plotted as in Fig. \ref{fig1}, but for
a smaller  orbital velocity  range, such that the peaks due to the excitation
of the stellar modes are excluded.
}
\label{fig4}
\end{figure}

\section{Concluding Remarks}\label{sec6}
In this review we have shown how to estimate the characteristic  features of the 
gravitational signals emitted by some  interesting astrophysical sources.

Chances to detect these waves crucially  depend  on the
detailed knowledge of the waveform; indeed
the standard technique used to extract  a non-continuous  signal from the noisy data 
of a detector is  the matched filtering technique, whose performances are
very sensitive to a mismatching of the parameters: for this reason 
the templates must be as accurate as  possible.
Since the most interesting signals are emitted in spacetime regions where the non linearity
of the gravitational interaction plays an important role, the ultimate template
would be that obtained by integrating the fully non linear equations of 
General Relativity.
However,  fully nonlinear simulations of phenomena like the gravitational collapse
or the coalescence of binary systems are, at present, at a preliminary
stage, although major progresses in the field are underway.
Thus, for the time being we need to rely on the results obtained  by approximate
methods, like the quadrupole formalisms and the
perturbative approach, to mention  those we have described in this paper. 
There are many things that we learn from these approaches.
Consider for instance the evolution of a binary system: the quadrupole formalism, applied
to point masses, allows to determine with high accuracy  the gravitational signal emitted 
when the two bodies are far apart, the frequencies and the expected amplitudes.
Including radiation reaction effects and using the adiabatic approximation,
we can predict how the orbit evolves, and how much time it takes to become
circular.  However, if the system is composed by a solar type star and a close planet (or
a brown dwarf companion), new phenomena may occur, like the resonant excitation of the g-modes
of the star, which can be accounted for  only if  the  internal structure 
of the star is included in the picture; 
in this case, the perturbative approach provides the tools needed to evaluate 
how much energy is  radiated in gravitational waves because of the stellar
oscillations,  at which frequencies, and
for howlong can a system be in a resonant condition.

If the binary system is formed by very compact objects, like neutron stars or  black holes, the
quadrupole approach works up to much
higher frequencies, since in this case 
the masses  can really be treated, to a large extent,
as point masses.
During the latest phases of the inspiralling,  let say
during the 10-20 cicles that preceed the merging when the signal frequency
is of the order of 900-1000 Hz,
the non linearity of the interaction produces deviations from the quadrupole signal which
can be estimated by introducing post-newtonian
corrections at  higher order of $(v/c)$ 
(see e.g. \cite{damouriyersathia} and references quoted there).
In the standard post-newtonian (PN) approach,
the coalescing bodies are still considered as point masses, and this is good enough for black
holes. However, in the case of NS-binaries
the fundamental mode of the stars may be excited in the same region, and 
this excitation may produce
corrections to the quadrupole signal 
that are comparable or higher than the PN corrections, and depend on the
internal structure of the stars;  as we have shown in Sec. \ref{sec5},
these phenomena can be studied again by a perturbative
approach, which has been shown to work remarkably well, even when the two stars are at a
distance as short as  three stellar radii \cite{tutti2}.

An accurate detection of the signal emitted during
the last few  cicles before coalescence may enlighten unresolved issues related
to the internal composition of neutron stars and to 
the state of matter at supernuclear density.
These effects, however will not be seen by the first generation of interferometers, for
which the most likely source to be detected is the coalescence of BH-BH binaries
with masses of the order of $20-30 ~M_\odot$
\cite{grishuck}. But there is hope for the future: advanced detectors are beeing proposed
like EURO, a european project now under a feasibility study,
that should be extremely sensitive at high frequency,  and would allow to study 
the processes we have described.

\section*{Acknowledgments}
The work on relativistic stellar perturbations described in this lecture
has been supported by the EU Programme 'Improving the Human
Research Potential and the Socio-Economic Knowledge Base' (Research
Training Network Contract HPRN-CT-2000-00137).


\begin{thebibliography}{99}
\bibitem{web}
For details of the interferometric  experiments see the web sites:
virgo.infn.it,\ ligo.caltech.edu,\
geo600.uni-hannover.de,\ tamago.ntk.nao.ac.jp,\ lisa.jpl.nasa.gov
\end{thebibliography}
\end{document}